\documentclass[a4paper,11pt]{article}
\pdfoutput=1  
\usepackage{jcappub}
\usepackage[T1]{fontenc} 
\usepackage{natbib}

\newcommand{\HCd}{\mathcal{H}}

\def\HCdt0{\tilde{\HCd}_{0}}

\newcommand{\dd}{\,\mathrm{d}}

\title{\boldmath Dark energy and dark matter unification from dynamical space time: Observational constraints and cosmological implications}

\author[a,1]{Fotios K. Anagnostopoulos,\note{Corresponding author.}}
\author[b,c]{David Benisty}
\author[d,f]{Spyros Basilakos}
\author[b,c,e]{Eduardo I. Guendelman}

\affiliation[a]{National and Kapodistrian University of Athens, Physics Department, Panepistimioupoli Zografou, 15772, Athens, Greece}
\affiliation[b]{Physics Department, Ben-Gurion University of the Negev, Beer-Sheva 84105, Israel}
\affiliation[c]{Frankfurt Institute for Advanced Studies (FIAS), Ruth-Moufang-Strasse~1, 60438 Frankfurt am Main, Germany}
\affiliation[d]{Academy of Athens, Research Center for Astronomy and Applied Mathematics, Soranou Efesiou 4, 11527, Athens, Greece}
\affiliation[e]{Bahamas Advanced Study Institute and Conferences, 4A Ocean Heights, Hill View Circle, Stella Maris, Long Island, The Bahamas}
\affiliation[f]{National Observatory of Athens, V. Paulou and I. Metaxa 15236, Penteli, Greece}
\emailAdd{fotis-anagnostopoulos@hotmail.com}
\emailAdd{benidav@post.bgu.ac.il}
\emailAdd{svasil@academyofathens.gr}
\emailAdd{guendel@bgu.ac.il}
\date{\today}

\abstract{A recently proposed Dynamical Space-time Cosmology (DSC) that unifies dark energy and dark matter is studied. The general action of this scenario includes a Lagrange multiplier, which is coupled to the energy momentum tensor 
and a scalar field which is different from quintessence. 
First for various types of potentials we implement
a critical point analysis and we find solutions which lead to
cosmic acceleration and under certain conditions to 
stable late-time attractors. Then the DSC cosmology is confronted with the latest cosmological data from low-redshift probes, namely measurements of the Hubble parameter and standard candles (Pantheon SnIa, Quasi-stellar objects). Performing an overall likelihood analysis and using the appropriate information criteria we find that the explored DSC models are in very good agreement with the data.
We also find that one of the DSC models  shows a small but non-zero deviation from $\Lambda$ cosmology, nevertheless 
the confidence level is close to $\sim 1.5\sigma$.}
\begin{document}
\maketitle
\flushbottom
\section{Introduction}
Almost twenty years after the observational evidence of cosmic acceleration the cause of this phenomenon, labeled 
as "dark energy" (hereafter DE), remains an open question which challenges the foundations of theoretical physics: The cosmological constant problem - why there is a large
disagreement between the vacuum expectation value of the
energy momentum tensor which comes from quantum field
theory and the observable value of dark energy density \cite{Weinberg:1988cp}-\cite{Lombriser:2019jia}. 
The simplest model of DE is the so called $\Lambda$CDM model that contains non-relativistic matter and cosmological constant. Although, this model fits accurately the present cosmological data it suffers from two fundamental problems, namely the tiny value of the cosmological constant and also the 
coincidence problem, \cite{Bull:2015stt}. 
Furthermore, there is criticism 
on the conceptual foundations of the current view 
of the cosmos, in a sense that there are too 
many {\it {ad hoc}} hypotheses (e. g dark Energy, dark matter) needed for 
"explaining the phenomena", e. g \cite{Merritt:2017xeh}. 

The main argument of 
the latter article is that whenever a scientific theory encounters difficulties in explaining phenomena, adding auxiliary hypotheses within the body of the theory, is considered bad practice. This is so as it could lead to \emph{non-falsifiable} theories, \cite{Popper}. The aforementioned criticism is not founded upon physical considerations so it can not be used right away to construct a cosmological model. However, it motivates the development of alternative cosmological models, which could 
provide a more natural description of the so called dark sector.

Unification between dark energy and dark matter from an action principle was obtained from scalar fields \cite{Scherrer:2004au}-\cite{Fadragas:2013ina}, by a complex scalar field \cite{Arbey:2006it} or other models \cite{Chen:2008ft}-\cite{Leon:2013qh} including Galileon cosmology \cite{Leon:2012mt} or Telleparallel gravity \cite{Kofinas:2014aka}-\cite{Skugoreva:2014ena}. Beyond those approaches, a unification of Dark Energy and Dark Matter using a new measure of integration (the so-called Two Measure Theories) has been formulated \cite{Guendelman:2015jii}-\cite{Guendelman:2016kwj}. A diffusive interacting of dark energy and dark matter models was introduced in \cite{Koutsoumbas:2017fxp}-\cite{Haba:2016swv} and it has been found that diffusive interacting dark energy - dark matter models can be formulated in the context of an action principle based on a generalization of those Two Measures Theories in the context of quintessential scalar fields \cite{Benisty:2017eqh}-\cite{Benisty:2017rbw}, although these models are not equivalent to the previous diffusive interacting dark energy - dark matter models.
In order to overcome the coincidence problem, Gao, Kunz, Liddle and 
Parkinson \cite{Gao:2009me}
suggested a unification of dark energy and dark matter resulting 
from a single scalar field. 
Unlike usual quintessence model, here 
the scalar field behaves either as dark matter or dark energy. 
Within this framework the unified picture of dark sector 
introduces a number of modifications in the equations 
of motion of the aforementioned scalar field. 
Recently, a Lagrangian formulation was introduced in 
\cite{Benisty:2018qed} (see also \cite{Guendelman:2009ck}) 
toward building the so called Dynamical Space-time Cosmological (DSC) model. 
In this scenario the gravitational field is 
described not only by the metric tensor but also by
a Lagrange multiplier that is coupled to the energy 
momentum tensor, a scalar field potential and another 
potential that describes the interactions between 
DE and DM.
The scalar field $\phi$ plays an important role in the 
description of the dynamics, since the kinetic 
term of $\phi$ behaves as DM and the 
potential is responsible for DE. 
Therefore, the DSC model provides an elegant alternative 
in describing the DM and DE dominated eras respectively.

In the current paper we attempt to continue our previous work 
of \cite{Benisty:2018qed} in the sense that we study both dynamically and 
observationally the DSC scenario
for a large family of potentials. 
Specifically, the manuscript is organized as follows. 
In Sec. II we briefly present the theoretical framework 
of the Dynamical Space-time Cosmological model 
and provide the basic cosmological equations.
In section III we use a dynamical analysis by studying 
the critical points of the field equations in the dimensionless variables
for a large family of potentials. 
In Sec. IV we provide the likelihood analysis 
and the observational data sets that we 
utilize in order to constraint the free parameters of the DSC model and we 
compare it with the $\Lambda$CDM model. Finally, in Sec. V we 
draw our conclusions.

\section{Dynamical Space-time Cosmology}
In previous publications two of us (Benisty and Guendelman, 
\cite{Benisty:2018qed}) proposed 
the Dynamical Space-time Cosmological model (DSC)
via a space time vector field, and demonstrated 
the behavior of this scenario toward unifying the dark sector.
In this section we briefly present the main features  
of the DSC model based on first principles.
The action that describes the gravitational field equations and 
unifies the dark sector was first introduced by Benisty and Guendelman \cite{Benisty:2018qed}:
\begin{equation}
\mathcal{S}=\int \sqrt{-g} \, \left[\frac{1}{2\kappa^2}R+\chi_{\mu;\nu} T^{\mu\nu}_{(\phi)} - \frac{1}{2}g^{\alpha\beta} \phi_{,\alpha}\phi_{,\beta} - V(\phi)\right]\, d^4x,
\end{equation}
where $\phi$ is the scalar field and $R$ is the Ricci scalar and $\kappa^2 = 8\pi G = 1$. 
The vector field $\chi_\mu$ is the so called dynamical space 
time vector, hence the corresponding covariant derivative is 
$ \chi_{\mu;\nu}=\partial_{\nu}\chi_{\mu}-\Gamma_{\ \mu\nu}^{\lambda}\chi_{\lambda}$, where
$\Gamma_{\ \mu\nu}^{\lambda} $ is the Christoffel symbol. 
In this context, $T^{\mu\nu}_{(\phi)}$ denotes the stress energy 
tensor which was first introduced by Gao and colleagues \cite{Gao:2009me} 
\begin{equation}
T^{\mu\nu}_{(\phi)} 
= -\frac{1}{2} \phi^{,\mu} \phi^{,\nu} + U(\phi) g^{\mu \nu}  .
\end{equation}
Obviously the action integral contains two different potentials, namely 
$U(\phi)$ which is coupled to the stress energy momentum tensor $T^{\mu\nu}_{(\phi)}$, and $V(\phi)$ 
which is minimally coupled into the action. Moreover, the action depends on three different quantities: 
the scalar field $\phi$ the dynamical space time vector $\chi_\mu$ and the metric $g_{\mu\nu}$.
\subsection{Equations of motion}
There are 3 independent variations for this theory. The first variation is with respect to the dynamical spacetime vector field $\chi_\mu$ which yields the conservation of the energy momentum tensor $T^{\mu\nu}_{(\phi)}$:
\begin{equation}
\nabla_{\mu} T^{\mu\nu}_{(\phi)} = 0.
\end{equation}
The second variation with respect to the scalar field $\phi$ gives a non-conserved current: 
\begin{subequations}\label{current}
\begin{equation}
\chi^\lambda_{;\lambda} U'(\phi) - V'(\phi) = \nabla_\mu j^{\mu} 
\end{equation}
\begin{equation}
j^{\mu}  = \frac{1}{2}\phi_{,\nu} (\chi^{\mu;\nu}+\chi^{\nu;\mu}) + \phi^{,\mu}
\end{equation}
\end{subequations}
and the derivatives of the potentials are the source of this current. For constant potentials the source term becomes zero and we get a covariant conservation of the current.

Lastly, varying the action integral with respect to the metric, we derive the gravitational field equations 
\begin{equation}\label{setv}
\begin{split}
\frac{1}{\kappa^2}G^{\mu\nu}= g^{\mu\nu} \left(\frac{1}{2}\phi_{,\alpha} \phi^{,\alpha} +  V(\phi)
+\frac{1}{2}\chi^{\alpha;\beta}\phi_{,\alpha} \phi_{,\beta}+\chi^{\lambda}\phi_{,\lambda}U'(\phi)\right) 
\\ 
- \frac{1}{2} \phi^{,\mu} \left[\left(\chi^{\lambda}_{;\lambda}+2\right) \phi^{,\nu} + \chi^{\lambda;\nu}\phi_{,\lambda} + \chi^\lambda \phi^{,\nu}_{;\lambda}\right] - \frac{1}{2}\left(\chi^{\lambda}\phi^{,\mu}_{;\lambda}\phi^{,\nu}+\chi^{\lambda;\mu}\phi_{,\lambda}\phi^{,\nu}\right).
\end{split}
\end{equation}
\subsection{Homogeneous Solution}
The (FLRW) Friedman-Lemaitre-Robertson-Walker ansatz is the standard model of cosmology dynamics based on the assumption of a homogeneous and isotropic universe at any point, commonly referred to as the cosmological principle. The symmetry considerations lead to the FLRW metric
\begin{equation}\label{eq:robwalk}
\dd s^{2}=\dd t^{2}-a^{2}(t)\left[\frac{\dd r^{2}}{1-Kr^{2}}+r^{2}\left(\dd\theta^{2}+\sin^{2}\theta\dd\phi^{2}\right)\right].
\end{equation}
Herein, $a(t)$ defines the dimensionless cosmological expansion (scale) factor. For simplicity we consider a homogeneous scalar field $\phi=\phi(t)$, while 
the dynamical vector $\chi_\mu$ is given by the following 
formula $\chi_\mu = \left(\chi_0,0,0,0\right)$, where $\chi_0$ is also just a function of time.

Varying the action  
with respect to the dynamical space time vector field $\chi_\mu$ 
we obtain the modified ''Klein-Gordon'' equation 
\begin{equation}\label{1frw}
\ddot{\phi}+\frac{3}{2}H\dot{\phi}+U'(\phi)=0,
\end{equation}
where the prime denotes derivative with respect to $\phi$. Compared with the equivalent equation which comes from quintessence model, this model gives a different and smaller friction term, as compared to the canonical scalar field. Therefore for increasing redshift, the densities for the scalar field will increase slower than in the standard quintessence.

The second variation, for homogeneous $\phi$ and 
$\chi_\mu = \left(\chi_0(t),0,0,0\right)$ Eq.(\ref{current})
becomes
\begin{equation}
j^{\mu}  = (\dot \phi (1-\dot{\chi}_0),0,0,0),
\end{equation}
hence for FRWL metric we obatin 
\begin{equation}
\begin{split}
j^{\mu}_{;\mu}  = \frac{1}{\sqrt{-g}} \partial_{\mu} \left(\sqrt{-g} j^{\mu} \right)  = \frac{1}{a^3} \partial_{\mu} \left(a^3 j^{0} \right) = -\ddot{\phi}(\dot{\chi}_0 - 1)+\dot{\phi}\left[3H(\dot{\chi}_0 - 1) - \ddot{\chi}_0\right]
\end{split}
\end{equation}
and the source term yields:
\begin{equation}
\begin{split}
\chi^\lambda_{;\lambda} U'(\phi) - V'(\phi) = U'(\phi) \frac{1}{\sqrt{-g}} \partial_{\mu} \left(\sqrt{-g} \chi^{\mu} \right) - V'(\phi) \\= -U'(\phi) \frac{1}{a^3} \partial_{\mu} \left(a^3 \chi^{0} \right) - V'(\phi)= U'(\phi)\left[\dot{\chi}_0+3H\chi_0\right] + V'(\phi).
\end{split}
\end{equation}
Therefore, the equation of motion takes the form:
\begin{equation}\label{chi}
\begin{split}
\ddot{\phi}(\dot{\chi}_0 - 1)+\dot{\phi}\left[3H(\dot{\chi}_0 - 1) + \ddot{\chi}_0\right] = U'(\phi)\left[\dot{\chi}_0+3H\chi_0\right] - V'(\phi).
\end{split}
\end{equation}
For the spatially homogeneous cosmological case the energy density and the pressure of the scalar field read:
\begin{subequations}
\begin{equation}
\rho=\dot{\phi}^2 (\dot{\chi}_0(1-\frac{3}{2}\mathcal{H})-\frac{1}{2}) + V(\phi) -\dot{\phi}\dot{\chi}_0 (U'(\phi)+\ddot{\phi})
\end{equation}
\begin{equation}
p = \frac{1}{2}\dot{\phi}^2(\dot{\chi}_0-1) - V(\phi) -\chi_0 \dot{\phi} U'(\phi)
\end{equation}
\end{subequations}
Comparing the stress energy tensor with equations (\ref{1frw}), we provide the functional forms of the energy density and pressure respectively:
\begin{equation}
\label{setvv}
\rho = \left(\dot{\chi}_0-\frac{1}{2}\right)\dot{\phi}^2 + V(\phi)
\end{equation}
\begin{equation}
\label{setvv1}
p = \frac{1}{2}\dot{\phi}^2(\dot{\chi}_0-1) - V(\phi) -\chi_0 \dot{\phi} U'(\phi).
\end{equation}
Unlike usual DE models, quintessence and the 
like, here the vector field $\chi_{0}$ and the potential $U(\phi)$ modify
the density and the pressure of the cosmic fluid. In order to proceed with the analysis we need to know the forms of $U(\phi)$ and $V(\phi)$. Below, we consider special forms of the potentials and study the performance of the models at the expansion level.

\subsubsection{Coupled Constant potential into the Lagrange multiplier}
Similar to \cite{Benisty:2018qed} we consider DSC models for which 
the potential $U(\phi)$ that is coupled to 
the stress energy momentum tensor $T^{\mu\nu}_{(\phi)}$ is constant 
\begin{equation}
U(\phi)= Const.
\end{equation}
The general study of varying $U(\phi)$ will appear in a forthcoming 
paper. Substituting the potential into Eq.(\ref{1frw}), using the definition of 
$H={\dot a}/a$  and performing the integration we find 
\begin{equation}
\label{phi}
\dot{\phi}^2 = \frac{2 C_{m0}}{a^3},
\end{equation}
where $C_{m0}>0$ is the integration constant which 
can be viewed as the effective dark matter energy density parameter.
Introducing the new variable 
\begin{equation}\label{delta}
\delta = \dot{\chi}_0 - 1
\end{equation}
equations Eqs. (\ref{chi}), (\ref{setvv}) and (\ref{setvv1})   
become
\begin{equation}
\label{phisec}
\dot{\phi}\left(\dot{\delta}+\frac{3}{2}H\delta\right) = - V'(\phi) .
\end{equation}
In this context, the energy density and the pressure of the scalar field
are given by 
\begin{subequations}
\begin{equation}\label{density}
\rho = \left(\delta+\frac{1}{2}\right)\dot{\phi}^2 + V(\phi),
\end{equation}
\begin{equation}\label{pressure}
p = \frac{\delta}{2}  \dot\phi^2-V(\phi).
\end{equation}
\end{subequations}

Furthermore, if we assume $V(\phi)=\Lambda=Const.$ then  
the solution of Eq. (\ref{phisec}) is 
\begin{equation}\label{chidot}
\delta = \frac{1}{2}\xi \, a^{-3/2}, 
\end{equation}
where $\xi$ is a dimensionless integration constant and
hence with the aid of (\ref{phi}) we obtain
 
\begin{subequations}
\begin{equation}
\label{constAllrho}
\rho = \Lambda +\frac{\xi C_{m 0}}{a^{9/2}} +\frac{C_{m0}}{a^3} 
\end{equation}
\begin{equation}
\label{constAllpress}
p = -\Lambda+ \frac{\xi C_{m 0}}{2\, a^{9/2}},
\end{equation}
\end{subequations} 

In such a case it is trivial to show that the Hubble parameter is given by
\begin{equation}
\label{costPotHubblerate}
H(z) = H_{0}\left[\Omega_{\Lambda } +\Omega_{m0}(1+z)^3 +\Omega_{\xi 0}(1+z)^{9/2} \right]^{1/2},
\end{equation}
where $\Omega_{\xi 0}=\xi \Omega_{m0}$ and $H_{0}$ is the Hubble constant, while we normalize the first Friedmann equation by the critical density $\rho_c = 3H_0^2$: $\Omega_\Lambda := \Lambda/ \rho_c$, $\Omega_{m0} := C_{m0}/\rho_c$.
The current model can be seen as an approximation of the general $U$,$V$ potentials, namely close to the present era where the potentials vary slowly with time. Therefore, the Hubble expansion Eq.(\ref{costPotHubblerate}) resembles that 
of the general case only in the late universe. Moreover, in the case of $\xi<0$ the latter situation holds for $z<z_{\rm max}$, where
$z_{\rm max} \simeq (-\Omega_{m0}/\Omega_{\xi 0})^{2/3}-1$. 
For a barotropic cosmic fluid whose the corresponding equation of state parameter is given by $w_{i}=p_{i}/\rho_{i}$ one can easily 
recognize three "dark fluids", namely 
cosmological constant [$w_{\Lambda} = -1$, $V(\phi)=Const.$], dark 
matter ($w_{m} = 0$), 
and another fluid with $w_{\xi} = 1/2$.
 
Notice that in the case of $\chi_0 = t$, from Eq. \ref{delta} and \ref{chidot} we get:
\begin{equation}
\delta = \xi = 0.
\end{equation}
Therefore, $\Lambda$CDM model is precisely obtained from Eq. \ref{constAllrho} and \ref{constAllpress}, namely
\begin{subequations}
\begin{equation}
\label{constAllrho2}
\rho = \Lambda +\frac{C_{m0}}{a^3} 
\end{equation}
\begin{equation}
\label{constAllpress2}
p = -\Lambda,
\end{equation}
\end{subequations} 
Lastly, it is interesting to mention that for $\Omega_{\xi 0} \neq 0$ we can get bouncing solutions as discussed in 
\cite{Benisty:2018qed}. 
\subsubsection{Dynamical DM-DE}
Here let us concentrate on a more general situation for which $U(\phi) = Const.$ and $V=V(\phi)$. 
Within this framework, the combination of equations (\ref{phi}), (\ref{phisec}), (\ref{density}) and 
(\ref{pressure}) provide 
\begin{subequations}
\label{systemForH}
\begin{equation}
\label{deltadz}
\frac{d \delta}{dz} = \frac{V'(\phi)}{(z+1)^{5/2}\sqrt[]{C_{m0}}H(\phi,\delta)} + \frac{3\delta}{2(z+1)}
\end{equation}
\begin{equation}
\label{phidz}
\frac{d \phi}{dz} = -\frac{\sqrt{2C_{m0}(z+1)}}{H(\phi,\delta)}
\end{equation}
with the Hubble parameter 
\begin{equation}
\label{systemForHH}
H(\phi,\delta) = H_0 \left[(2\delta +1)\Omega_{m0}\left(z+1\right)^3+\Omega_{DE}(\phi) \right]^{1/2} ,
\end{equation}
\end{subequations}
where $\Omega_{DE}(\phi) = V(\phi)/\rho_{c}$ and $z=a^{-1}-1$ is the redshift.
Therefore, in order to derive the evolution of the Hubble parameter 
we need to solve the system of equations 
(\ref{deltadz}), (\ref{phidz}). 

Suppose that we know the functional form of the potential $V(\phi)$. 
First we evaluate Eq.(\ref{systemForHH}) at $z=0$ which means that
$(2\delta(z=0)+1)\Omega_{m0}+ \Omega_{DE}(\phi(z=0))=1$. 
Second, the fact that $V(\phi) \sim \Lambda $ 
prior to the present time together with the cosmic sum
$\Omega_{DE,0}+\Omega_{m0}=1$ imply $\delta(z=0) = 0$, hence 
the form of $V(\phi)$ obeys 
$V(\phi) = \Lambda f(\phi)$, where $f(\phi) = 1$ at $z=0$. Concerning the types of $V(\phi)$ potentials involved in the present analysis,  we consider the following three cases: exponential with
$V(\phi) = V_0 e^{-\beta\phi}$, cosine with $V(\phi) = V_0 \cos{\beta\phi}$ and thawing potential with $V(\phi) = V_0 e^{-\alpha \phi} (1+\beta \phi)$, \cite{Clemson:2008ua}.
This family of models has $\Lambda$CDM as an asymptotic solution.  
Notice that the initial condition for $\phi$ is chosen appropriately 
to be compliant with the aforementioned constrain, that is $\phi (z=0) = 0$. 
Once steps (i) and (ii) are accomplished, we numerically 
solve the system (\ref{deltadz}), (\ref{phidz}). 
\section{Dynamical system method}
In this section we provide a dynamical analysis 
by studying the fixed points of the field equations, 
so that we can investigate 
the various phases of the current cosmological models. 
Specifically, for a general potential $V(\phi)$ 
we introduce the new dimensionless variables 
\begin{equation}\label{def}
x = \frac{\dot{\phi}}{\sqrt{6}H} , 
\quad y = \frac{\sqrt{V(\phi)}}{\sqrt{3}H}, \quad 
z = -\frac{V'(\phi)}{V(\phi)}.
\end{equation}
In the new system of variables the field equations form an 
autonomous system which is given by
\begin{subequations}
\begin{equation}\label{asm1}
x' = -\frac{3}{4} x \left(x^2+3 y^2-1\right)
\end{equation}
\begin{equation}\label{asm2}
y' = -\frac{1}{4} y \left(3 x^2 + 9 y^2- 9+2 \sqrt{6} x z\right)
\end{equation}
\begin{equation}\label{asm3}
z' = -\sqrt{6} z^2 x \left( \Gamma - 1 \right), 
\end{equation}
\end{subequations}
where 
\begin{equation}
\Gamma = \frac{V(\phi)V''(\phi)}{V'(\phi)^2}.
\end{equation}
These are the basic variables that we use for mapping the dynamical system. 
In this case 
the equation of motion with respect to the metric is written as:  
\begin{equation}\label{28}
(1+2\delta)x^2+y^2=1
\end{equation}
Notice that for $\chi_0 \sim t$, which means $\delta \sim 0 $, the phase plane of the system takes the form of a complete circle, where the points $(1,0)$ and $(0,1)$ correspond to dark matter and dark energy dominated eras respectively. 

Bellow we provide the results of the dynamical analysis for different types 
of potentials. The corresponding 
critical points of the system (\ref{asm1}), (\ref{asm2}) and 
(\ref{asm3}) are listed in Tables I, II and III. 
In all cases point A with coordinates $(0,0)$ is ruled 
out from the constrain (\ref{28}). 
\subsection{Exponential potential ($V_{1}$)}
We continue our work by using the exponential potential. In this case the new variable $z$ (see the last term in Eq.\ref{def}) becomes constant. 
The dynamical system includes four critical points, among which 
one point is stable. 
Point B with coordinates $(1,0)$ 
corresponds to the matter epoch and it is stable when 
$\beta>\sqrt{\frac{3}{2}}$. 
Point C with coordinates $(0,1)$  
describes the dark energy dominated era, while point D 
$\left(\beta>\sqrt{\frac{3}{2}}\right)$   
with coordinates $\left(\sqrt{\frac{3}{2}}\frac{1}{\beta},\sqrt{\frac{2\beta^{2}-3}{\sqrt{6}\beta}}\right)$ is unstable. 
\begin{table}[h]
  \begin{center}
  \label{tab:table1}
    \begin{tabular}{l|c|c|c|r} 
    \textbf{Name}  & \textbf{Stability} & \textbf{Universe} & \textbf{The point  $(x,y)$} \\
       \hline
      A &  unstable & - & $\left(0,0 \right)$ \\ 
      B &  stable for $\beta>\sqrt{\frac{3}{2}}$ & Dark Matter & $\left(1,0 \right)$ \\
      C & asymptotically stable & Dark Energy & $\left(0,1 \right)$ \\
      D ($\beta>\sqrt{\frac{3}{2}}$) & unstable saddle p. & unified DE-DM & $\left(\sqrt{\frac{3}{2}}\frac{1}{\beta},\frac{\sqrt{2 \beta ^2-3}}{\sqrt{6} \beta}\right)$\\
    \end{tabular}
\caption{Critical points for $V(\phi) \propto e^{-\beta \phi}$.}
  \end{center}
\end{table}
\subsection{Cosine potential ($V_{2}$)}
Now we proceed with the cosine potential $V\propto {\rm cos}(\beta \phi)$. 
Inserting this formula into $\Gamma$ we find 
\begin{equation}
z' = \sqrt{6} x (1 + z^2).
\end{equation}
Therefore, for the dynamical analysis we utilize 
the aforementioned equation together with Eqs. (\ref{asm1})-(\ref{asm2}). 
In this case we find three critical points which are not affected 
by $\beta$. 
As expected, points $B(1,0)$ and $C(0,1)$ describes the 
dark matter and dark energy dominates eras respectively. 
Here $B$ is always unstable, while $C$ is asymptotically stable.
\begin{table}[h]
  \begin{center}
  \label{tab:table2}
    \begin{tabular}{l|c|c|c|r} 
    \textbf{Name}  & \textbf{Stability} & \textbf{Universe} & \textbf{The point  $(x,y)$} \\
       \hline
      A &  unstable & - & $\left(0,0 \right)$ \\ 
      B &  unstable & Dark Matter & $\left(1,0 \right)$ \\
      C & asymptotically stable & Dark Energy & $\left(0,1 \right)$ \\
         \end{tabular}
         \caption{Critical points for the Cosine potential.}
  \end{center}
\end{table}
\subsection{Thawing potential ($V_{3}$)}
Using the thawing potential
$V(\phi) \propto e^{-\alpha \phi} (1+\beta \phi)$
(\ref{asm3}) becomes:
\begin{equation}
z' = \sqrt{6} x (z-\alpha)^2\;.
\end{equation}
In this case point $B(1,0)$ is unstable when $\alpha<\sqrt{3}$ 
and it is saddle for $\alpha>\sqrt{3}$. The dark energy point $C(0,1)$ 
is always stable. Lastly, point D
with coordinates $\left(\sqrt{\frac{3}{2\alpha}},
\sqrt{\frac{2\alpha-3}{6\alpha}}\right)$ is stable when 
$\alpha<\sqrt{\frac{5}{6}}$  
and it is saddle when $\alpha>\sqrt{\frac{5}{6}}$.  
\begin{table}[h]
  \begin{center}
  \label{tab:table3}
    \begin{tabular}{l|c|c|c|c|r} 
    \textbf{Name} & \textbf{Existence} & \textbf{Stability} & \textbf{Universe} &  \textbf{The point} $(x,y,z)$\\
       \hline
      A & all $\alpha$ & unstable & -  & $(0,0)$\\
      B &  all $\alpha$  & $\alpha<\sqrt{3}$ unstable, $\quad \alpha>\sqrt{3}$ saddle point. & Dark Matter & $(1,0,\alpha)$\\
      C &  all $\alpha$   & stable & Dark Energy  & $(0,1)$\\
      D & $\alpha>\sqrt{\frac{3}{2}}$  &$\alpha < \sqrt{\frac{5}{6}}$ stable focus, $\alpha > \sqrt{\frac{5}{6}}$ saddle & unified DE-DM & $\left(\sqrt{\frac{3}{2\alpha}},\sqrt{\frac{2\alpha-3}{6\alpha}},\alpha\right)$\\
    \end{tabular}
        \caption{Critical points for $V(\phi) \propto e^{-\alpha \phi} (1+\beta \phi)$}
  \end{center}
\end{table}
\begin{figure*}[t!]
\centering\includegraphics[width=1\textwidth]{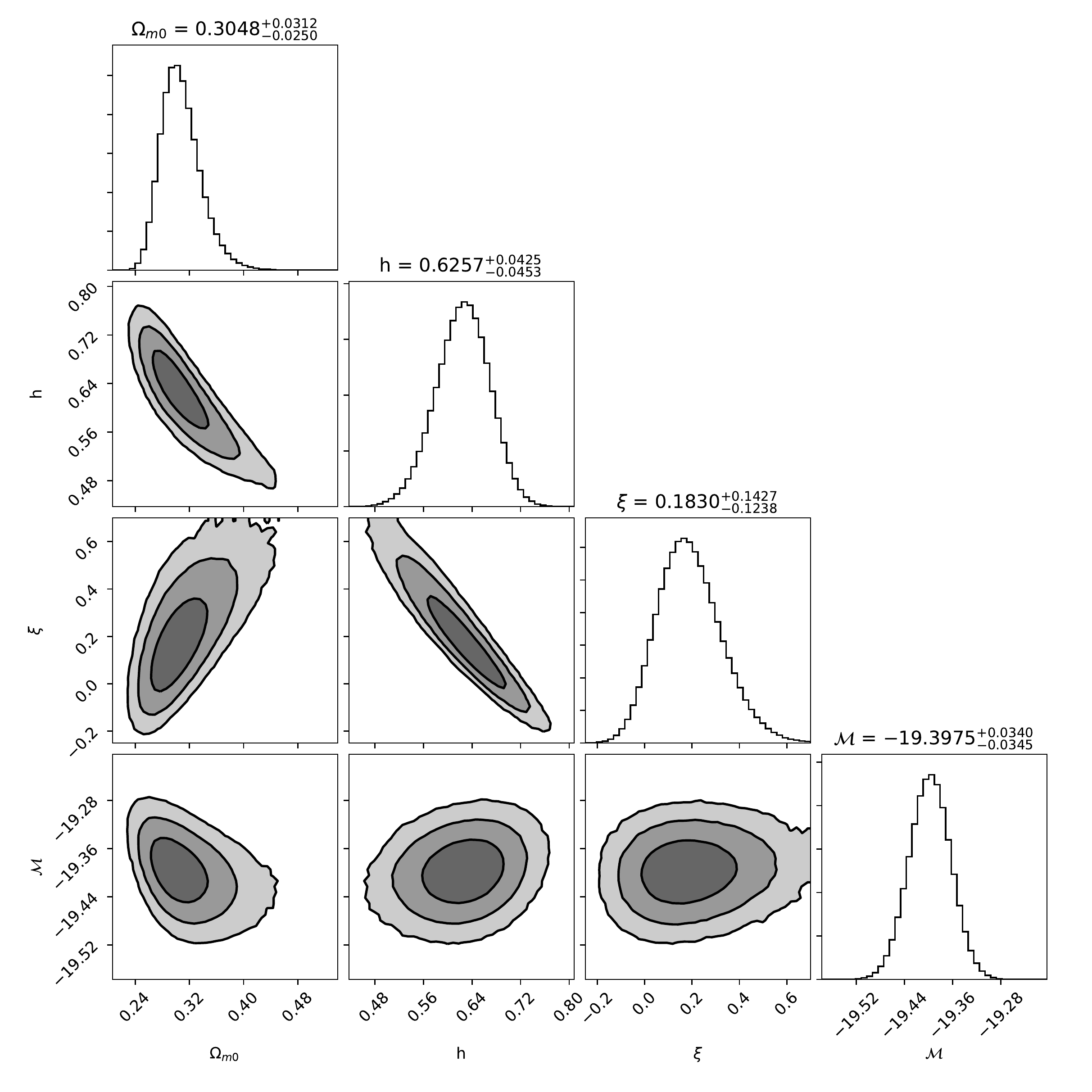}\\
\caption{Observational constraints of the DSC with V = Const. and $U(\phi)=$Const.}
\label{fig:Consts}
\end{figure*} 
\begin{figure*}[t!]
\centering\includegraphics[width=1\textwidth]{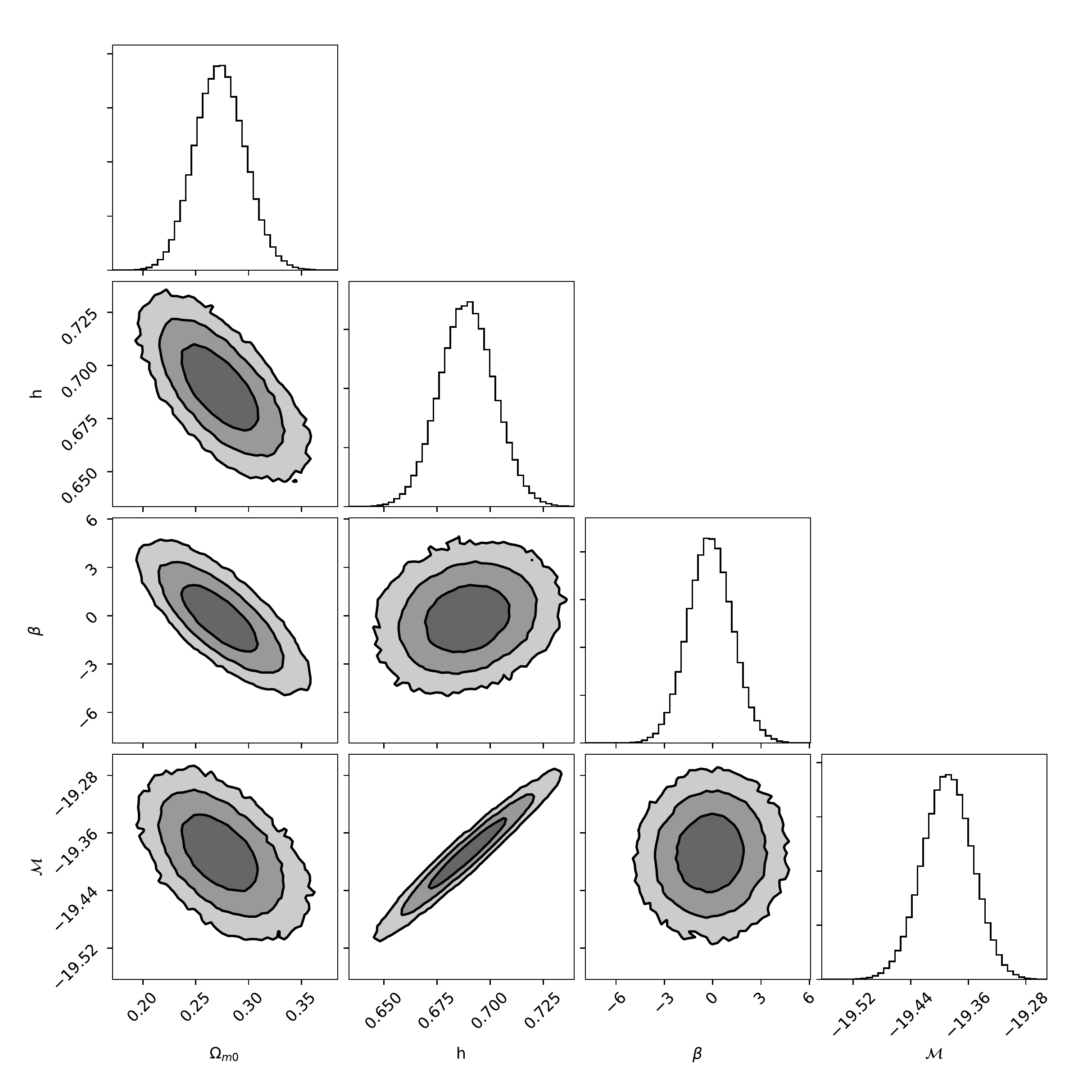}\\
\caption{Observational constraints of the exponential DSC, $V_1 \propto e^{-\beta \phi}$, while we have used $U(\phi)=$Const.}
\label{fig:V_1}
\end{figure*} 
\begin{figure*}[t!]
\centering\includegraphics[width=1\textwidth]{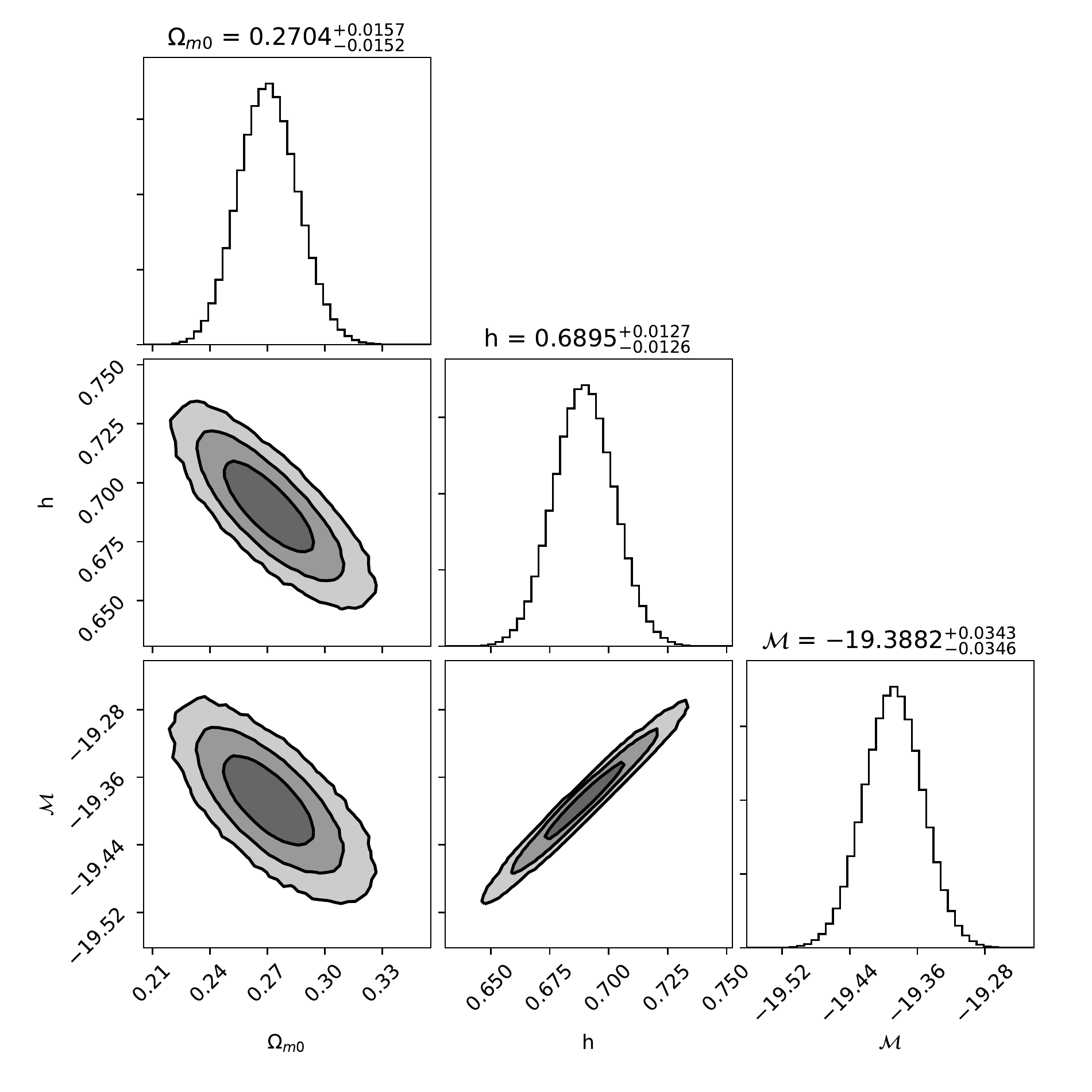}\\
\caption{Observational constraints of the cosine DSC, $V_2 \propto {\rm cos}(\beta \phi) $, while we have used $U(\phi)=$Const.}
\label{fig:V_2}
\end{figure*} 
\begin{figure*}[t!]
\centering\includegraphics[width=1\textwidth]{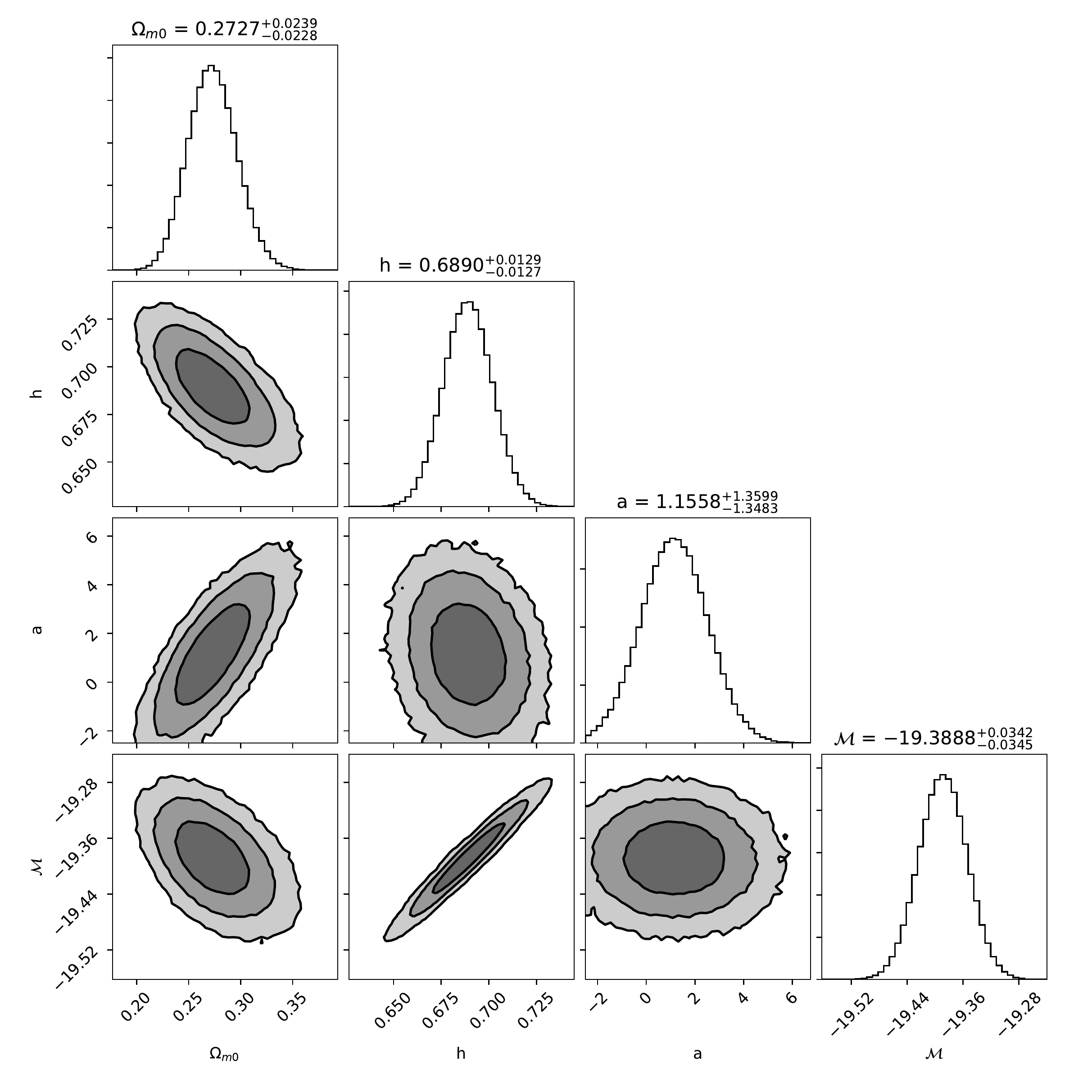}\\
\caption{Observational constraints of the thawing DSC, $V_3 \propto e^{-\alpha \phi} (1+\beta \phi) $, while we have used $U(\phi)=$Const.}
\label{fig:V_3}
\end{figure*} 

\section{Observational Constraints}
In the following we describe the observational data sets along 
with the relevant statistics in constraining 
the DSC models presented in Sect. III.
\subsubsection{Direct measurements of the Hubble expansion}
          We use the latest $H(z)$ data set compilation, that 
can be found in \cite{YuRatra2018}. 			
This set contains $N=36$
          measurements of the Hubble expansion in the 
following redshift range $0.07\leq z\leq 2.33$.
          Out of these, there are 5 measurements based on Baryonic Acoustic Oscillations (BAOs), while for the rest, the Hubble 					parameter
          is measured via the differential age of passive evolving galaxies. 
  
Here, the corresponding 
          $\chi^2_{H}$ function reads
          \begin{equation}
          \chi^{2}_{H}\left(\phi^{\nu}\right)={\bf \cal H}\,
          {\bf C}_{H,\text{cov}}^{-1}\,{\bf \cal H}^{T}\,,
          \end{equation}
          where ${\bf \cal H }=\{H_{1}-H_{0}E(z_{1},\phi^{\nu})\,,\,...\,,\,
          H_{N}-H_{0}E(z_{N},\phi^{\nu})\}$ and $H_{i}$ are the observed Hubble rates at redshift $z_{i}$ ($i=1,...,N$). Notice, that 
the statistical vector $\phi^{\nu}$  
contains the parameters that we want to fit. 
The matrix ${\bf C}$ denotes the covariance
          matrix. Further considerations regarding the statistical analysis and the corresponding covariance matrices can be found in Ref. \cite{AA} 
and references therein.

\subsubsection{Standard Candles}
          The second data-set that 
we include in our analysis is the binned Pantheon sample of Scolnic et al.
          \cite{Scolnic:2017caz} and the binned sample of Quasi-Stellar Objects (QSOs), \cite{RisalitiLusso:2015}.
          We would like to note the importance of using the 
Pantheon SnIa data along with those of QSOs. 
The latter allows to trace the cosmic history to the 
redshift range $0.07< z <6$. 
It is important to utilize alternative probes
at higher redshifts in order to verify the SnIa results and test
any possible evolution of the DE equation of state \cite{Plionis2011}.
          Following standard lines, the chi-square function 
of the standard candles is given by
          \begin{equation}
          \chi^{2}_{\text{s}}\left(\phi^{\nu}_{\text{s}}\right)={ \bf \mu}_{\text{s}}\,
          {\bf C}_{\text{s},\text{cov}}^{-1}\,{\bf \mu}_{\text{s}}^{T}\,,
          \end{equation}
          where
          ${\bf \mu}_{\text{s}}=\{\mu_{1}-\mu_{\text{th}}(z_{1},\phi^{\nu})\,,\,...\,,\,
          \mu_{N}-\mu_{\text{th}}(z_{N},\phi^{\nu})\}$ and the subscript $\text{s}$ denotes SnIa
          and QSOs. For the SnIa data the covariance matrix is not diagonal and the distance modulus is given as
          $\mu_{i} = \mu_{B,i}-\mathcal{M}$, where $\mu_{B,i}$ is the apparent magnitude at maximum in the rest frame for 
          redshift $z_{i}$ and $\mathcal{M}$ is treated as a universal free parameter, \cite{Scolnic:2017caz}, quantifying various observational uncertainties.
          It is apparent that $\mathcal{M}$ and $h$ parameters are intrinsically degenerate in
          the context of the Pantheon data set, so we can not extract any information regarding
          $H_{0}$ from SnIa data alone. In the case of QSOs, $\mu_{i}$ is the observed distant modulus
          at redshift $z_{i}$ and the covariance matrix is diagonal.In all cases, the theoretical
          form of the distance modulus reads
          \begin{equation}
          \mu_{\text{th}} = 5\log\left(\frac{D_{L}(z)}{\text{Mpc}}\right) + 25\,,
			\end{equation}
where
			\begin{equation}
D_L(z) = c(1+z)\int_{0}^{z}\frac{dx}{H(x,\theta^{\nu})}
\end{equation}
is the luminosity distance, for spatially flat FRWL cosmology.

        \subsubsection{Joint analysis and model selection}

        In order to perform a joint statistical analysis of $P$ cosmological probes (in our case $P=3$),
        we need to use the total likelihood function
        \begin{equation}
        \mathcal{L}_{\text{tot}}(\phi^{\psi}) = \prod_{p=1}^{P} \exp(-\chi^2_{p})\,.
        \end{equation}
        Consequently the  $\chi^2_{\text{tot}}$ expression is given by
        \begin{equation}
        \chi_{\text{tot}}^2 = \sum_{p=1}^{P}\chi^2_{P}\,,
        \end{equation}
        where the statistical vector has dimension $\psi$, which is the sum of the $\nu$
        parameters of the model at hand plus the number $\nu_{\text{hyp}}$ of hyper-parameters of the data
        sets used, that is $\psi = \nu + \nu_{\text{hyp}}$.The distinction between the hyper-parameters quantifying uncertainties in a data set and
        the free parameters of the cosmological model is purely conceptual. Regarding the problem of likelihood maximization, we use an affine-invariant Markov Chain Monte Carlo sampler, as described in Ref. \cite{AffineInvMCMC}. 
We utilize the open-source python package emcee, \cite{emcee}, using 1000 "walkers" and 1500 "states". The latter setup corresponds to $\sim 10^6$ calls of the total likelihood function. In each call, we need to numerically solve the system of Eqs. (\ref{systemForH}) for the redshift range [0.0,5.93] and also calculate the luminosity distance. This procedure became practical by optimizing critical parts of the calculations using C++ code from Ref. \cite{Mpampis}. The convergence of the MCMC algorithm is checked with auto-correlation time considerations.

\subsection{Statistical Results}
\begin{table*}[t!]
\tabcolsep 5.5pt
\vspace{1mm}
\begin{tabular}{ccccccc} \hline \hline
Model & $\Omega_{m0}$ & $h$ & $\alpha \ or \ \xi $  & $\beta$ & $\mathcal{M}$ &
$\chi_{\text{min}}^{2}$  \vspace{0.05cm}\\ \hline
\hline
 
  $V,U const.$ &$ 0.305_{-0.025}^{+0.031}    $ & $0.6257_{-0.0455}^{+0.0428} $ & $0.183_{-0.125}^{+0.143} $& - &$ -19.397_{-0.035}^{+0.034} $ & $ 84.114 $\\ 
 $V_1$ & $0.277_{-0.023}^{+0.024}$ & $0.6885_{-0.0128}^{+0.0130}$ & - & $-0.593_{-1.355}^{+1.367}$ & $-19.390_{-0.035}^{+0.034}$ & $88.100$\\ 
 $V_2 (cosine) $ & $ 0.270 \pm 0.015  $ & $0.6895_{-0.0127}^{+0.0128}  $ & $ - $& 1 &$-19.388
 \pm 0.035 $ & $ 87.954 $\\ 
$V_3$ & $0.273_{-0.023}^{+0.024}$ & $0.6890_{-0.0127}^{+0.0130} $ & $1.152_{-1.352}^{+1.370}  $& 1 &$-19.389 \pm 0.034$ & $87.942$\\ 
 
$\Lambda$CDM & $0.281^{+0.016}_{-0.015}$ & $0.686 \pm 0.013$& - & - & $-19.403
\pm 0.035$ & 85.700  \vspace{0.45cm}\\ 

\hline\hline
\end{tabular}
\caption[]{ Observational constraints and the
corresponding $\chi^{2}_{\rm min}$ for the considered cosmological models.
Notice that $\Omega_{\xi 0}=\xi\Omega_{m0}$.  
The concordance $\Lambda$ model is included for comparison.}

\label{tab:Results1}
\end{table*}

In order to test the performance of the cosmological models 
in fitting the data it is imperative to utilize 
the Akaike Information Criterion (AIC), \cite{Akaike1974}, and Bayesian Information Criterion (BIC), \cite{Schwarz1978}. 
The AIC criterion  is an asymptotically unbiased estimator of the Kullback-Leibler information,
        measuring the loss of information during the fit. Within the standard assumption of Gaussian errors, the
        AIC estimator is given by
        \cite{Ann2002}
        \begin{equation}
        \text{AIC}=-2\ln(\mathcal{L}_{\text{max}})+2\psi+
        \frac{2\psi(\psi+1)}{N_{\rm tot}-\psi-1}\,,
        \end{equation}
        where $\mathcal{L}_{\text{max}}$ is the maximum likelihood of the data set(s) under consideration
        and $N_{\rm tot}$ is the total number of data. It is easy to observe that for large $N_{\rm tot}$, this expression
        reduces to $\text{AIC}\simeq -2\ln(\mathcal{L}_{\text{max}})+2\psi$, that is the standard form of the AIC criterion. Following the previous point, it is considered good practice to use the modified AIC criterion in all cases, \cite{Liddle:2007fy}.
        
On the other hand, the BIC criterion is an estimator of the Bayesian evidence, (e. g \cite{Liddle:2007fy},\cite{Ann2002} and references therein), and is given as

\begin{equation}
\text{BIC} = -2\ln(\mathcal{L}_{\text{max}})+\psi \,{\rm log}(N_{\text{tot}})\,.
\end{equation}
The AIC and BIC criteria employ only the likelihood value at maximum. In principle, due to to the Bayesian nature of our analysis, the accuracy of the $\mathcal{L}_{max}$ is reduced, meaning that the AIC and BIC values are meant to be used just for illustrative purposes. In practice, however, by using long chains, we obtain $\mathcal{L}_{max}$ values with enough accuracy to use them 
in order to calculate AIC and BIC. 
Furthermore, we also compute the Deviance Information 
Criterion (DIC), that provides 
all the information obtained from the likelihood calls 
during the maximization procedure. 
The DIC estimator is defined as, (see \cite{Liddle:2007fy}, \cite{Spiegelhalter2002})

\begin{equation}
{\rm DIC} = D(\overline{\phi^\psi}) + 2C_{B},
\end{equation}
where $C_{B}$ is the so called 
Bayesian complexity that measures
the power of data to constrain the parameter space 
compared to the predictivity of the model
which is provided by the prior. In particular, 
$C_{B} = \overline{D(\phi^\omega)} - D(\overline{\phi^\omega})$,
where the overline denotes the usual mean value.
Also, $D(\phi^\omega)$ is the Bayesian Deviation, where
in our case it boils down to $D(\phi^\omega) = -2\ln(\mathcal{L(\phi^\omega)})$.

To proceed with the model selection we need to assign a "probability" to each model following the classical treatment of Jeffreys, \cite{Jeffreys}, that is by using the relative difference of the IC value for a number
        of models, $\Delta \text{IC}_{\text{model}}=\text{IC}_{\text{model}}-\text{IC}_{\text{min}}$,
        where the $\text{IC}_{\text{min}}$ is the minimum $\text{IC}$ value in the set of competing models. 
Following the notations of Ref. \cite{KassRaftery1995}, $\Delta\text{IC}\leq 2$, means that the model under consideration is statistically compatible with the ``best'' model, while the condition $2<\Delta\text{IC}<6$ indicates a middle tension between the two models and the condition $\Delta\text{IC}\geq 10$ suggests a strong tension.
\begin{table*}
\centering
\tabcolsep 5.5pt
\vspace{1mm}
\begin{tabular}{ccccccc} \hline \hline
Model & AIC & $\Delta$AIC & BIC &$\Delta$BIC & DIC & $\Delta$DIC
 \vspace{0.05cm}\\ \hline
\hline
$const$ & 92.535 & 0.582 & 102.535 & 3.020 & 88.567 & 0 \\ 
 $V_1$ &  96.521 & 4.569 & 106.5210 & 7.006 & 95.908 & 7.341 \\  
$V_2$ &  94.204  & 2.252 &  101.770 & 2.255 & 93.930 & 5.363 \\ 
$V_3$ &  96.363  & 4.411 &  106.363 & 6.848 & 95.805 & 7.238 \\ 
$\Lambda$CDM & 91.952 & 0 & 99.515 & 0 & 91.671 & 3.104 \vspace{0.45cm}\\ 


\hline\hline
\end{tabular}
\caption{The information criteria 
AIC, BIC and DIC for various cosmological models 
along with the corresponding differences
$\Delta\text{IC} \equiv \text{IC} - \text{IC}_{\text{min}}$.
\label{tab:Results2}}
\end{table*}

Utilizing the aforementioned likelihood analysis we summarize our statistical results in 
Table \ref{tab:Results1}.


For the model with constant potentials, we find  $\Omega_{m0} = 0.305_{-0.025}^{+0.031}, h = 0.6257_{-0.0455}^{+0.0428} , \xi = 0.183_{-0.125}^{+0.143} $ with $\chi^2_{min} = 84.114$. The relevant contours are present at Figure \ref{fig:Consts}. Interestingly, the $\xi = 0$ value which 
corresponds to the $\Lambda CDM$ limit is outside the $1\sigma$ area. 

Regarding the exponential potential ($V_1$), we find $\Omega_{m0} =0.277_{-0.023}^{+0.024}, \ h = 0.6885_{-0.0128}^{+0.0130}$, $ \beta = -0.593_{-1.355}^{+1.367} $ with 
$\chi^2_{min} = 88.100$ and the contours are in Figure \ref{fig:V_1}. 
Furthermore, the cosmological parameters for the cosine potential ($V_2$) are $\Omega_{m0} = 0.270 \pm 0.015, \ h = 0.6895^{+0.0128}_{-0.0127}$ and the relevant $\chi^2_{min} = 87.954$. The contour plots are presented in Figure \ref{fig:V_2}.
Lastly, for the potential $V_3$ we obtain the contours of Figure \ref{fig:V_3} and the parameter values: $\Omega_{m0} = 0.273_{-0.023}^{+0.024}, \ h = 0.6890_{-0.0127}^{+0.0130} , \ \alpha = 1.152_{-1.352}^{+1.370} $ and $\chi^2_{min} = 87.942$.  
In most of the cases, the best fit values of the 
matter energy density are in good 
agreement with those of Planck 2018, \cite{Aghanim:2018eyx}. 
Considering the result for the flat $\Lambda CDM$, we observe $1\sigma$ compatibility for the $V_{i}, \ i =1,2,3$ potentials, while the result for the cosmological model with constant $V,U$ potentials is within $2\sigma$ limits. 
It is important to note that for $V_2$ and $V_3$ potentials 
we set $\beta = 1 $. 
However, we have tested that the likelihood analysis 
provides very similar results for $\beta \sim{\cal O}(1)$.

\begin{figure*}[t!]
\centering\includegraphics[width=1\textwidth]{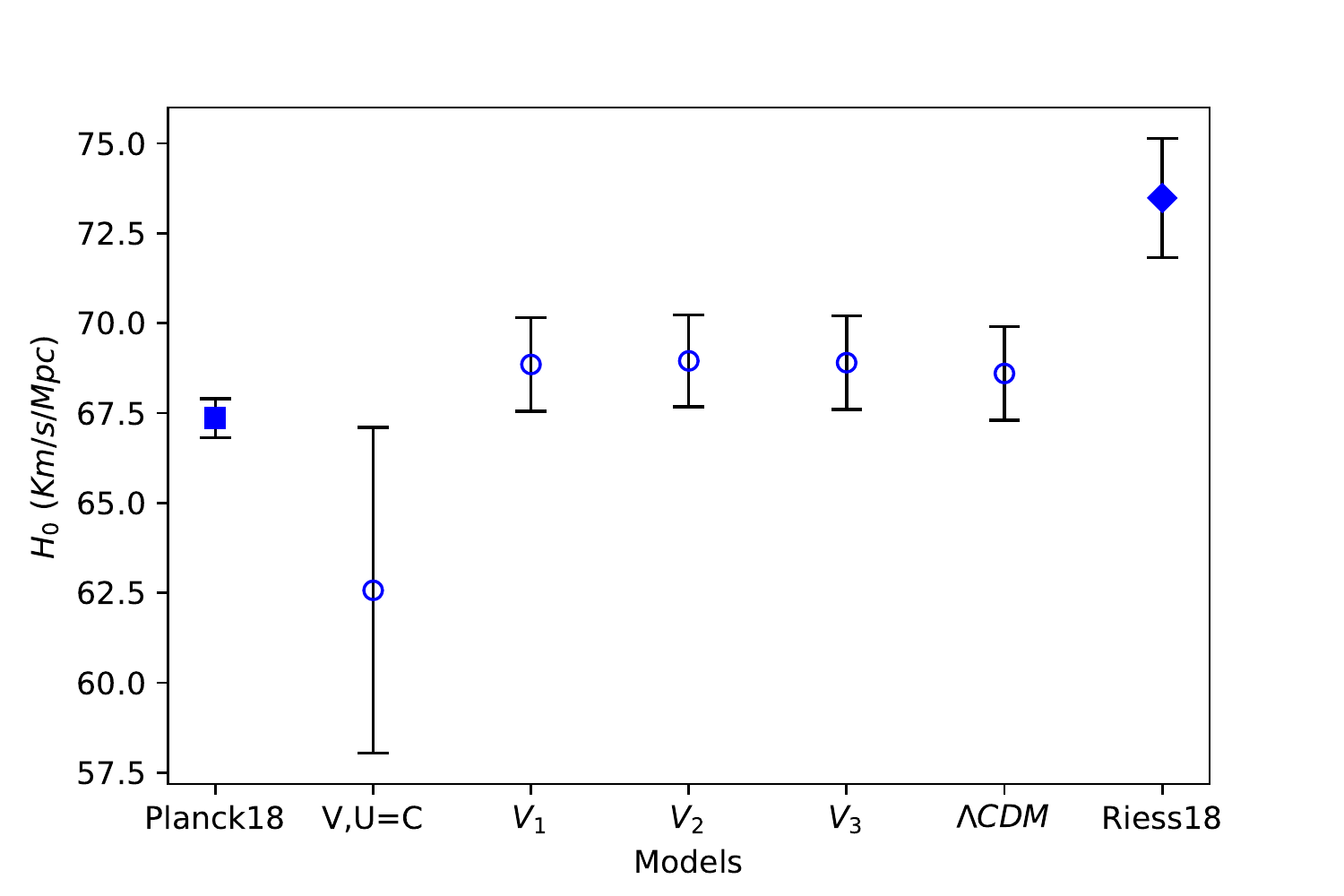}
\caption{A synopsis of our results regarding the Hubble constant problem. The labels 'Planck18' and 'Riess18' stand for the relevant results from Ref. \cite{Aghanim:2018eyx} and \cite{RiessHo} respectively.}
\label{fig4}
\end{figure*}

We deem interesting to discuss our results with respect 
to the Hubble constant problem, that is a $\sim 3.7\sigma$ discrepancy between the Hubble constant measured by Riess {\it{et al.} }, \cite{RiessHo}, ($H_{0}=73.48 \pm 1.66 Km/s/Mpc$) and the relevant value from Planck collaboration, ($H_{0}^{Planck} = 67.36 \pm 0.54$ Km/s/Mpc), \cite{Aghanim:2018eyx}, see Figure \ref{fig4}. 
Our results are in agreement (within $1\sigma$) 
with those provided by the team of Planck, 
while there is compatibility at
$\sim 2 \sigma$ level with Riess {\it{et al.}} results. However, the Hubble constant for the constant potentials case is significantly smaller from other relevant results, however due to the large error bar maintains $\sim 1 \sigma$ compatibility. As a consistency check we compare our results with the result from the model-independent assessment of the cosmic history obtained by Haridasu {\it{et al}}, \cite{Haridasu2018},  namely $H^{ind.}_{0} = 68.52 \pm 0.94$ and we report $1\sigma$ compatibility for $V_1,\ V_2, \ V_3$ potentials, while the $V,\ U$ constant potential is within $2\sigma$. 

Concerning AIC, BIC and DIC and we present the relevant values at the Table \ref{tab:Results2}. In the context of BIC, all models considered are in mild to strong tension with $\Lambda$CDM. 
As we used binned data sets, we do not anticipate that 
an information criterion with an explicit dependence 
from the dataset length could estimate reliably the 
relative quality of the fits. Further, the BIC 
criterion is just an asymptotic approximation that 
is valid while the dataset length tends to infinity. 
On the other hand, AIC criterion provides a somewhat different view. 
The model with constant potentials has $\Delta AIC \leq 2$, hence 
it is nearly indistinguishable from $\Lambda$CDM. The other 
models ($V_{i}, \ i = 1,2,3$) are in mild tension 
with $\Lambda$CDM since they have $2 < \Delta AIC < 6$. 
In the context of our Bayesian treatment, both AIC and BIC values could only serve indicative purposes, as they employ only the value of the likelihood at maximum and not the full set of likelihood values obtained during the sampling procedure. The most interesting observation comes from the DIC criterion, which seems to prefer the cosmological model with constant $(U,V)$ potentials over the concordance model, as $DIC_{\Lambda} > DIC_{constVU}$ and the relevant $\Delta DIC$ (>2) indicates that the difference is rather 
significant. However, as we mentioned before, the constant potentials model is an approximation of a more general cosmological model, valid for late universe only.
With respect to the other models under consideration, we observe mild - to - strong tension with each of them, with the $\Lambda CDM$ to be in the second place. A general ascertainment regarding the somewhat similar results of the physically different potentials in the free parameters (e.g matter energy density and Hubble constant) is that $\phi$ is very small at late universe, so any $V(\phi)$ is effectively $V(\phi) \sim \Lambda$ (where we have set $8\pi G=1$). This is what someone could naively foresee as the field $\phi$ changes very smoothly across the cosmic history. We expect that a study of the model using the CMB spectrum could discriminate between the different DSC models.

\section{Conclusions}
We explored a large family of cosmological models in the context of 
Dynamical Space-time Cosmology (DSC). 
This scenario unifies naturally the dark sector 
and it provides an elegant theoretical platform toward
describing the various phases of the cosmic expansion.
Initially, we performed a standard dynamical analysis and we found 
that under certain circumstances DSC model includes  
stable late-time attractors.  
Then we tested the class of DSC models against the latest 
observational data and we placed constraints 
on the corresponding free parameters.
In particular, our observational constraints 
regarding the Hubble constant are 
in agreement (within $\sim 1 \sigma$) with those of Planck 2018.
Moreover, our results are compatible at $\sim 2\sigma$ level 
with the $H_0$ measurement obtained from Cepheids.

Using the most popular information criteria we found cases 
for which the DSC model is statistically equivalent 
with that of $\Lambda$CDM and thus it can be viewed
as a viable cosmological alternative. On top of that we found that 
one of the DSC models, that with $V(\phi)=Const.$ and 
$U(\phi)=Const.$, shows a small but non-zero deviation from $\Lambda$CDM, where
the confidence level is close to $\sim 1.5\sigma$. Also, we explicitly checked that our $V_{1,2,3}$ models are able to pass the BBN constraints (see Appendix). 
We argued that the theoretical formulation of 
Ref. \cite{Benisty:2018qed} could provide competitive 
cosmological models and thus it deserves further consideration. 
Finally, in a forthcoming paper we attempt to investigate
DSC at the perturbation level 
for the general case of 
potentials $U(\phi)$. This will allow us to modify CAMB and 
thus to confront Dynamical Space-time Cosmology 
to the Cosmic Microwave Background (CMB) power spectrum from Planck.

\acknowledgments
FA wishes to thank Charalambos Kioses for a number of very interesting discussions and also for vast help at optimization of the code for this project. This article is supported by COST Action CA15117 "Cosmology and Astrophysics Network for Theoretical Advances and Training Action" (CANTATA) of the COST (European Cooperation in Science and Technology) and the actions CA18108, CA16104. SB would like to acknowledge support by the Research Center for Astronomy of the Academy of Athens in the context of the program "Tracing  the  Cosmic  Acceleration".

\appendix

\section{Big Bang Nucleosynthesis (BBN) within DSC}
In the appendix we check the various DSC models against 
BBN. 
Of course, the complete analysis of this aspect is out of scope 
of the present study. However, we explicitly checked the 
compatibility of DSC within the 
standard BBN using the average bound on the possible variation of the BBN
speed-up factor. 
The latter is defined as the ratio of the expansion rate predicted in a given model
versus that of the $\Lambda CDM$ model at the BBN epoch, namely 
$z_{\rm BBN} \sim 10^9$.
Specifically, using the best fit values (see Table 4) 
regarding the cosmological parameters ($\theta_{i}^{\nu}$)
we check the validity of the following inequality, (i.e \cite{Sola:2016jky} and references therein)
$$ 
100\% \times  \frac{(H_{\Lambda CDM}(z_{\rm BBN},\Omega_{m0},h) - H_{i}(z_{\rm BBN},\theta^{\nu}_{i}))^2}{H_{\Lambda CDM}(z_{\rm BBN},\Omega_{m0},h)^2} < 10\%.
$$
Notice that $i = 1, 2, 3$ correspond to 
exponential, cosine and thawing potentials respectively (see section 3).
We verify that the latter potentials satisfy the above restriction, hence 
BBN is safe in the context of DSC cosmology.
Concerning the concordance $\Lambda$CDM model we have used 
that provided by the Planck team \cite{Aghanim:2018eyx}.


\bibliographystyle{apsrev4-1}

\end{document}